# Methods for Joint Imaging and RNA-seq Data Analysis

Junhai Jiang[1], Nan Lin[1], Shicheng Guo[1,2], Jinyun Chen[3] and Momiao Xiong[1,*]

[1]Human Genetics Center, Division of Biostatistics, The University of Texas School of Public Health, Houston, TX 77030, USA

[2]State Key Laboratory of Genetic Engineering and Ministry of Education, Key Laboratory of Contemporary Anthropology, Collaborative Innovation Center for Genetics and Development, School of Life Sciences and Institutes of Biomedical Sciences, Fudan University, Shanghai 200433, China

[3]Department of Epidemiology, Division of OVP, Cancer Prevention and Population Sciences, The University of Texas MD Anderson Cancer Center, Houston, TX 77030, USA

[*] Address for correspondence and reprints: Dr. Momiao Xiong, Human Genetics Center, The University of Texas Health Science Center at Houston, P.O. Box 20186, Houston, Texas 77225, (Phone): 713-500-9894, (Fax): 713-500-0900, E-mail: Momiao.Xiong@uth.tmc.edu

**Abstract**

Emerging integrative analysis of genomic and anatomical imaging data which has not been well developed, provides invaluable information for the holistic discovery of the genomic structure of disease and has the potential to open a new avenue for discovering novel disease susceptibility genes which cannot be identified if they are analyzed separately. A key issue to the success of imaging and genomic data analysis is how to reduce their dimensions. Most previous methods for imaging information extraction and RNA-seq data reduction do not explore imaging spatial information and often ignore gene expression variation at genomic positional level. To overcome these limitations, we extend functional principle component analysis from one dimension to two dimension (2DFPCA) for representing imaging data and develop a multiple functional linear model (MFLM) in which functional principal scores of images are taken as multiple quantitative traits and RNA-seq profile across a gene is taken as a function predictor for assessing the association of gene expression with images. The developed method has been applied to image and RNA-seq data of ovarian cancer and KIRC studies. We identified 24 and 84 genes whose expressions were associated with imaging variations in ovarian cancer and KIRC studies, respectively. Our results showed that many significantly associated genes with images were not differentially expressed, but revealed their morphological and metabolic functions. The results also demonstrated that the peaks of the estimated regression coefficient function in the MFLM often allowed the discovery of splicing sites and multiple isoform of gene expressions.

**Significance**

Despite imaging-genetics shows great promise as a powerful tool for dissecting genomic structures of complex diseases; to date, very few imaging and RNA-seq analyses have been performed. We present a novel model for the integrative analysis of imaging and RNA-seq data and offer a new paradigm for RNA-seq data analysis. The results show that imaging and RNA-seq analysis can detect cancer susceptibility genes that are not differentially expressed. Surprisingly, most image associated genes display alternative splicing, which change the protein structures and cell morphologies. The results also demonstrate that the peaks of regression coefficient functions in the model were located in the splicing sites. Integrative imaging and RNA-seq analysis opens a new avenue for identifying disease causing genes.

\body

There is increasing consensus that imaging measures show closer associations with genomic variants and the penetrance of an individual genomic variant is expected to be higher at the imaging level than at the clinical diagnostic and outcome level. Imaging measures as an endophenotype have a higher power to identify genomic variants that significantly contribute to the development of diseases (1, 2). Integrated genomic and imaging data analysis is a new powerful approach used to uncover the individual variability and mechanism of disease development (3). Both imaging and genomics generate a huge amount of data that present critical bottlenecks in their analysis. Despite its great success, integrative analysis of unprecedented high dimensional imaging and genomic data faces great conceptual and computational challenges (4).

A key issue to the success of imaging and genomic data analysis is how to reduce dimensions of both imaging and genomic data. Previously investigated methods for imaging information extraction include single region-of-interest (ROI) methods, voxelwise approaches, principal component analysis (PCA), singular value decomposition, self-organizing Map (SOM) and multidimensional scaling (MDS) (5). However, these multivariate dimension reduction methods do not explore imaging spatial information. They take the set of spectral images as an unordered set of high dimensional pixels (6). Spatial information is very important for image cluster and classification analysis. To overcome limitations of multivariate dimension reduction and to utilize spatial information of the image signal, we extend the widely used one dimensional functional principal component analysis (FPCA) (7) to high dimensional FPCA to extract imaging signals.

The traditional methods for assessing the relationship between gene expressions measured by microarray and phenotypes are linear regressions (8, 9). However, the rapidly developed next-generation sequencing (NGS) technologies have become the platform of choice for gene expression profiling. RNA-seq for expression profiling offers a comprehensive picture of the transcriptome, with less background noise and a wider dynamic range of expression (10). Unlike microarrays for measuring gene expression, RNA-seq provides multiple layers of resolutions and transcriptome complexity: the expression at exon, SNP, and positional level, splicing, transcription start sites, polyadenylation sites, post-transcriptional RNA editing across the entire gene, and isoform and allele-specific expression (11). The current linear regression for modeling association of gene expressions with phenotypes quantifies the expression level of a gene/transcript by a single number that summarizes all the reads mapped to that gene/transcript. A single number measuring gene expression level ignores gene expression variation at the genomic positions. Therefore, linear regression is appropriate for microarray expression data, but may not be good for RNA-seq data.

To overcome these limitations, we propose a multiple functional linear model (MFLM) in which functional principal component scores of images are taken as multiple quantitative traits and RNA-seq profile across a gene is taken as a function predictor for assessing the association of gene expression with imaging signals which can take gene splicing and expression variation at genomic positional level into account.

**Results**

To evaluate its performance, the proposed MFLM for integrative imaging and RNA-seq data analysis was applied to images and RNA-seq datasets of ovarian cancer (OV) and kidney renal clear cell carcinoma (KIRC) which were downloaded from the TCGA datasets. The ovarian cancer dataset consists of 231 tumor tissue samples with histology images and RNA-seq profiles of 16,598 genes (after quality control). The KIRC dataset consists of 188 (121 tumor and 67 normal tissue samples) with histology images and RNA-seq profiles of 16,775 genes (after quality control). RNA-seq data were created by Illumina HiSeq 2000 PE paired-end RNA sequencing. More detailed information can be downloaded from TCGA website (http://cancergenome.nih.gov/).

**FPCA for Imaging Signal Extraction**

In our study, we compared our two dimensional FPCA with the traditional PCA by capturing space variation of image signals. To evaluate the performance of the two methods on image compression, we compared the original histology images and reconstructed images by two dimensional FPCA and by PCA. The result clearly demonstrated that the reconstructed images by FPCA were much closer to the original images than that by PCA (Fig. S1). In addition, 90.3% of the total imaging variation could be explained by top 30 functional principal components, while only 63.6% of the total imaging variation was explained by top 30 traditional principal components. Therefore, two dimensional FPCA is a better and more authentic image compression algorithm for image signal capturing than traditional PCA with minimal loss of information and fewer principal components usage.

**Behavior of the MFLM for Integrative Analysis of RNA-seq and Imaging Data**

In the process of integrative analysis of RNA-seq and image data, histology image data were compressed with our proposed two dimensional FPCA, and the FPC scores were taken as phenotypes. We considered gene expression values at single-base resolution and represented the expression profile of a gene by a functional curve, called a "gene expression function". We used the Karhunen-Loeve decomposition (7) to decompose the random gene expression function into orthogonal FPCs. The multiple FPC scores for imaging signal extraction were regressed on the FPC scores that were obtained from decomposition of the gene expression functions. In other words, we proposed to use MFLM for integrative analysis of RNA-seq and imaging data (Materials and Methods). Two FPCs that accounted for 81.7% and 88.6% of variation of imaging signals for ovarian and KIRC, respectively, were selected as phenotypes. The number of selected FPCs for the RNA-seq which accounted for 95% of the variation of gene expression ranges from 2 to 60. P-values for declaring significant association after applying the Bonferroni correction for multiple tests in ovarian cancer and KIRC analysis were $3.012 \times 10^{-6}$ and $2.98 \times 10^{-6}$, respectively. To indirectly examine the validity of MFLM for assessing the association of gene expression with the histology images, we plotted a QQ plot of the test in the MFLM (Fig.1). The QQ plots clearly showed that the false positive rates of the MFLM for detection of the association of gene expression with histology images in both ovarian cancer and KIRC studies were controlled.

**MFLM for Integrative Analysis of RNA-seq and Histology Images**

Three statistical methods: MFLM with FPC scores as phenotypes, MFLM with image descriptors (12) as phenotypes and multivariate regression model with FPC score as phenotypes and a single gene expression value (level 3 in TCGA datasets) as a regressor were applied to the ovarian cancer and KIRC datasets. For the ovarian cancer dataset, MFLM with FPC scores as phenotypes, MFLM with image descriptors and multivariate regression identified 24, 2 and 0 genes whose expressions were associated with image signals, respectively. Similarly, for the KIRC dataset, MFLM with FPC scores as phenotypes, MFLM with image descriptors and multivariate linear model (MLM) identified 84, 6 and 1 genes whose expressions were associated with image signals, respectively. The results were summarized in Tables 1 and 2.

Several remarkable features from these results were observed. First, the P-values calculated from the MFLM with FPC scores as phenotypes were much smaller than that calculated from the MFLM with image descriptors as phenotypes. Two methods assumed the same functional linear model (FLM) for RNA-seq data, but with a different approach to imaging signal reduction. The FPCA can reduce the dimensions of the imaging data more substantially than the traditional image descriptors. Therefore, the degrees of the test statistic in the MFLM with FPC score as phenotypes were much smaller than that in the MFLM with descriptors as phenotypes, which lead to the smaller P-values of the tests in the MFLM with the FPC scores as phenotypes. Second, we observed very few significant associations of the gene expression in the MLM. The MLM used the same FPCA for imaging data reduction, but model the gene expression level in a gene as a single value. The results demonstrated that the widely used single value representation of the expression level in the gene overlooked the expression variation across the gene, which led to large P-values of the tests. Third, expressions of genes which were associated with the imaging signal may or may not be differentially expressed (Table S1, Fig. S2). In other words, significant association of gene expression with imaging signals can provide additional information which differential expressions cannot offer. For example, genes *NOTCH1*, *ARHGEF11* and *BRD4* that were associated with imaging signals, but not differentially expressed between tumor and normal tissues were reported to regulates interactions between physically adjacent cells and induce G2/M arrest and triggers apoptosis in renal cell carcinoma (13), associated with kidney injury in the Dahl salt-sensitive rat (14) and kidney disease (15). Fourth, the MFLM with FPC scores as phenotypes could identify associated genes that showed alternative splicing expression pattern.

To illustrate this, we presented average expression of microtubule associated tumor suppressor 1 (*MTUS1*) in the KIRC study (Fig. 2). So far, seven isoforms of *MTUS1* have been discovered. We observed from Fig. 2 that a higher expression level in exon 1, exon 2 and exon 15 in the normal samples than that in tumor samples, and alternatively spliced transcript variations encoding different isoforms between tumor and normal samples were substantial. To our surprise, in Fig. 2 we also observed a remarkable feature that locations of the two peaks of the regression coefficient function $\beta(s)$ were close to the splicing sites at the genomic positions 17613470 and 17554765. This might indicate that splicing sites affect the tissue structure variation which was measured by imaging signals. *MTUS1* is interacted with microtubules to control cellular architecture and organize microtubule arrays. Express variation of *MTUS1* influences variation in microtubule structure, which in turn causes variation of histology images. Disruption of microtubule-dependent processes is involved in cancer development and metastasis (16). Fifth, imaging data convey relatively closer association with the disease than traditional phenotypes (17). The genes significantly associated with imaging will have profound implication in cellular function and disease development.

In the ovarian cancer study, among the 24 significantly associated genes with histology images, protein tyrosine phosphatase receptor type G (*PTPRG*) that regulate a variety of cellular processes including cell growth, differentiation, mitotic cycle, and oncogenic transformation, is a functional tumor suppressor gene and involved in ovarian tumorigenesis (18, 19). Cytoplasmic polyadenylation element binding protein 3 (*CPEB3*) that controls cell cycle progression, regulates senescence, establishes cell polarity, and promotes tumorigenesis and metastasis (20), plays a role in ovarian cancer development (21). In the KIRC study, integrin, alpha 9 (ITGA9) that participates in regulation of myotube formation (22), is reported to be involved in renal carcinomas (23), *NOTCH1* that regulates interactions between physically adjacent cells, is reported to trigger apoptosis in renal cell carcinoma (24), and Rho guanine nucleotide exchange factor (GEF) 11 (ARHGEF11) whose expression induces the reorganization of the actin cytoskeleton and the formation of membrane ruffling and filopodia, is associated with kidney injury (25) and key regulators of tumorigenesis (26).

**Image Associated Gene Form Protein-Protein Interaction Networks**

A large proportion of genes whose expression variation was associated with imaging signal variation formed protein-protein interaction networks (Fig. 3). In the ovarian cancer study, proteins of 30 out of 130 significantly associated genes identified by false discovery rate are interacted with each other to form a network. Hub gene *SETDB1*, encoding a histone methyltransferase in the network, is an oncogene and is involved in the development of several cancers (27). Another hub gene *Glul* that catalyzes the synthesis of glutamine from glutamate and ammonia is involved in cell proliferation, inhibition of apoptosis, and cell signaling, plays key roles in several cancers (28). We also observed from the KIRC study that proteins of 28 out of the 84 significantly associated genes with imaging signals are interacted to form a network, in which 10 genes are differentially expressed between tumor and normal tissues. A hub gene *REV3L* with 11 degrees in the network is the catalytic subunit of DNA translesion synthesis polymerase $\zeta$. It involves a variety of DNA-damaging, genome stability, cytotoxicity, and resistance to chemotherapeutic agents. Surprisingly, although *REV3L* is not differentially expressed, it is reported to be associated with lung, breast, colon cancers and gliomas (29-31). The interacted genes *KCNN3, ANKRD17, BRD4, NOTCH1, SMAD2, ZMIZ1, UFD1L,* and *MINK1* are associated with various cancers (32-38). Most of these genes are not differentially expressed, but are involved in the formation of cell and tissue structures. Their gene expression variations cause imaging signal variations and are thereby captured by integrative RNA-seq and imaging analysis.

**Image Associated Genes and Alternative Splicing**

We performed FPCA on RNA-seq profiles of each image associated gene and obtained their FPC scores in the ovarian cancer and KIRC studies. Then, we used a hierarchical algorithm to cluster genes based on their FPC scores. The results were shown in Fig. 4 and Fig. S3. We used DAVID (the Database for Annotation, Visualization and Integrated Discovery) Bioinformatics Resources (39), to extract biological features/meaning. DAVID bioinformatics gene function annotation analysis showed that most image associated genes play important roles in alternative splicing (Fig. 2 and Fig. S2). We observed that the genes with the similar patterns of alternative splicing sites are grouped together (Fig. S4). There is increasing consensus that alternative splicing may affect large and conservative regions of the protein structures and often leads to changes in cell morphologies and phenotypes such as actin cytoskeleton remodeling, regulation of cell-cell junction formation and regulation of cell migrations (40, 41). Variations in alternative splicing of gene expression lead to variations in cell morphologies and phenotypes, thus influencing variations of imaging measures of the cells. This opens a new pathway to cancer development and progression.

**Discussion**

The current major focus on RNA-seq data analysis is to identify differentially expressed genes (42) and major paradigm of RNA-seq data analysis is to test differences in gene expression level that is measured by a single value of summarizing statistic. However, there is increasingly recognition that the differential expression feature of genes may not be a unique source to cause disease. Changes in cell morphologies and motility can also influence development and progression of diseases. In this paper we have presented a MFLM with FPC scores of imaging measures as phenotypes for the integrative analysis of imaging and RNA-seq data and offered a new alternative paradigm for RNA-seq data analysis. We have also shifted the paradigm of RNA-seq data analysis from the single value representation of gene expression to the random function representation of RNA-seq profile which takes gene expression variation at the genomic positional level into account. Our study has made several remarkable findings.

The first finding is that imaging and RNA-seq analysis can detect cancer susceptibility genes that are not differentially expressed. Changes in cell morphologies, motility and phenotypes play important roles in the development and progression of the cancer. Genes causing these changes may not be differentially expressed between tumor and normal tissue samples and hence cannot be detected by gene differential expression analysis. The integrative analysis of imaging and RNA-seq data opens a new avenue for identifying cancer causing genes.

The second finding is that the function feature of image associated genes is alternative splicing. Surprisingly, we found that the peaks of regression coefficient functions in the MFLM of imaging and RNA-seq data analysis were located in the splicing sites. Alternative splicing often changes the protein structures, cell morphologies and phenotypes (40, 41). These changes generate variation of histology images of tumor tissues, which in turn provide information for discovery of image associated genes.

The third finding is that the widely used single value representation of the expression level in the gene overlooks the expression variation at the genomic positional level across the gene and hence has great limitations to identify image associated genes.

As demonstrated in the real data analysis, the MFLM showed great promise as a tool for integrative analysis of imaging and RNA-seq data. However, to date, very few integrative analyses of imaging and RNA-seq data have been performed. The results presented in this paper are among the first such studies and hence are considered preliminary. The number of selected orthogonal basis functions in the expansion of RNA-seq function will influence the performance of the integrative analysis of imaging and RNA-seq data. Genome-wide imaging and RNA-seq data analysis still poses great challenges. The main purpose of this paper is to stimulate discussion on the optimal strategies for genome-wide imaging and RNA-seq data analysis.

**Methods**

**Two dimensional functional principal component analysis**

One dimensional functional principal component analysis (FPCA) has been well developed (7). Now we extend one dimensional FPCA to two dimensional FPCA. Consider a two dimensional region. Let $s$ and $t$ denote coordinates in the $s$ axis and $t$ axis, respectively. Let $x(s,t)$ be a centered image signal located at $s$ and $t$ of the region. The signal $x(s,t)$ is a function of locations $s$ and $t$.

Consider a linear combination of functional values:

$$f = \int_S \int_T \beta(s,t) x(s,t) ds dt,$$

where $\beta(s,t)$ is a weight function. To capture the variations in the random functions, we chose weight function $\beta(s,t)$ to maximize the variance of $f$, which, after imposing a constraint to make the solution unique, leads to the following optimization problem:

$$\max \quad \int_S \int_T \int_S \int_T \beta(s_1,t_1) R(s_1,t_1,s_2,t_2) \beta(s_2,t_2) ds_1 dt_1 ds_2 t_2$$

$$\text{s.t.} \quad \int_S \int_T \beta^2(s,t) ds dt = 1. \tag{1}$$

where $R(s_1,t_1,s_2,t_2) = \text{cov}(x(s_1,t_1), x(s_2,t_2))$ is the covariance function of the image signal function $x(s,t)$. By variation calculus (43), we obtain the eigenequation as a solution to the optimization problem (1):

$$\int_S \int_T R(s_1,t_1,s_2,t_2)\beta_j(s_2,t_2)ds_2 dt_2 = \lambda \beta_j(s_1,t_1), j=1,2,...,J \qquad [2]$$

for an appropriate eigenvalue $\lambda$, where $\beta_j(s,t)$ is an eigenfunction. The random functions $x_i(s,t)$ can be expanded in terms of eigenfunctions as

$$x_i(t,s) = \sum_{j=1}^{K} \xi_{ij}\beta_j(s,t), i=1,...,N, \qquad [3]$$

where $\xi_{ij} = \int_S \int_T x_i(t,s)\beta_j(s,t)ds dt$, $i=1,...,N, j=1,...,J$ are FPC scores (Supplementary note 1).

**Multivariate Functional Linear Model for Integrative Analysis of Imaging and RNA-seq Data**

We take $K$ FPC scores as $K$ quantitative traits. Assume that $n$ individuals are sampled. Let $y_{ik}, k=1,2,...,K,$ be $K$ trait values of the $i$-th individual. Consider a genomic region $[a, b]$. Let $x_i(t)$ be a RNA-seq profile, the number of reads as a function of the genomic position $t$, of the $i$-th individual defined in the regions $[a, b]$. The multivariate functional linear model (MFLM) for integrative analysis of imaging and RNA-seq data can be defined as

$$y_{ik} = \alpha_{0k} + \int_T \alpha_k(t)x_i(t)dt + \varepsilon_{ik}, \qquad [4]$$

where $\alpha_{0k}$ is an overall mean, $\alpha_k(t)$ are a regression coefficient function for the $k$-th trait, $k=1,...,K$, $\varepsilon_{ik}$ are independent and identically distributed normal variables with mean of zero and covariance matrix $\Sigma$.

We assume that both trait values and RNA-seq profiles are centered. The RNA-seq profiles $x_i(t)$ are expanded in terms of the orthonormal basis function as:

$$x_i(t) = \sum_{j=1}^{J} \xi_{ij}\phi_j(t), \qquad [5]$$

where $\phi_j(t)$ are sequences of the orthonormal basis functions. Substituting equation (5) into equation (4), we obtain

$$y_{ik} = \sum_{j=1}^{J} \xi_{ij}\alpha_{kj} + \varepsilon_{ik}, i = 1,...,n, k = 1,...,K, \qquad [6]$$

where $\alpha_{kj} = \int_T \alpha_k(t)\phi_j(t)dt$. The parameters $\alpha_{kj}$ are referred to as genetic additive effect scores for the $k$-th trait.

Equation (6) can be rewritten in a matrix form:

$$Y = \xi\alpha + \varepsilon.$$

The standard least square estimators of $\alpha$ and the variance covariance matrix $\Sigma$ are given by

$$\hat{\alpha} = (\xi^T\xi)^{-1}\xi^T(Y - \bar{Y}),$$

$$\hat{\Sigma} = \frac{1}{n}(Y - \xi\hat{\alpha})^T(Y - \xi\hat{\alpha}).$$

Denote the matrix $(\xi^T\xi)^{-1}\xi^T$ by $A$. Then, the estimator of the parameter $\alpha$ is given by

$$\hat{\alpha} = A(Y - \bar{Y}).$$

The variance-covariance matrix of the estimator of the parameter $\alpha$ is given by

$$\begin{aligned}\Lambda = \text{var}(vec(\hat{\alpha})) &= (I_k \otimes A)(\Sigma \otimes I_n)(I_k \otimes A^T) \\ &= \Sigma \otimes (AA^T)\end{aligned} \qquad [7]$$

An essential problem in the QTL analysis or in the integrative analysis of imaging and RNA-seq data is to test the association of a gene with imaging phenotype. Formally, we investigate the problem of testing the following hypothesis:

$\alpha_k(t) = 0, \forall t \in [a,b], k = 1,..., K,$

which is equivalent to testing the hypothesis:

$H_0 : \alpha = 0$.

Define the test statistic for testing the association of a gene with $K$ quantitative traits as

$$T = vec(\hat{\alpha})^T \Lambda^{-1} vec(\hat{\alpha}). \qquad [8]$$

Let $r = \text{rank}(\Lambda)$.

Then, under the null hypothesis $H_0 : \alpha = 0$, $T$ is asymptotically distributed as a central $\chi^2_{(KJ)}$ or $\chi^2_{(r)}$ distribution if $J$ components are taken in the expansion equation (5) (Supplementary note 2).

**Acknowledgments**

The project described was supported by grants 1R01AR057120–01 and 1R01HL106034-01 from the National Institutes of Health and NHLBI, respectively. The authors wish to acknowledge the contributions of the research institutions, study investigators, field staff and study participants in creating the TCGA datasets for biomedical research.

**Figure Legends**

**Fig. 1.**

**A.** QQ plot for the KIRC dataset.

**B.** QQ plot for the ovarian cancer dataset.

**Fig. 2.**

**A.** Number of reads as a function of the genomic position of gene MTUS1.

**B.** Regression coefficient function of gene MTUS1 in the MFLM.

**Fig. 3.**

**A.** Proteins of 30 out of 130 significantly associated genes identified by false discovery rate are interacted with each other to form a network in the ovarian cancer study where genes in yellow color were differentially expressed between tumor and normal tissues and dotted vertical lines denote location of splicing sites.

**B.** Proteins of 28 out of 84 significantly associated genes are interacted each other to form a network in the KIRC study.

**Fig. 4.** Clusters of image associated genes in the ovarian cancer study by $k$-means clustering algorithms.

**Supplementary Figures**

**Fig, S1.**

**A.** Original ovarian cancer histology image.

**B.** Reconstruction of the ovarian cancer histology images from 30 FPC scores (explain 90.3% of the variations).

**C.** Reconstruction of the ovarian cancer histology images from 30 PC scores (explain 63.6% of the variations).

**D.** Original KIRC histology image.

**E.** Reconstruction of the kidney cancer histology image from 30 FPC scores (explain 93.6% of the variations).

**F.** Reconstruction of the kidney cancer histology images from 30 PC scores (explain 57.4% of the variations).

**Fig. S2**. RNA-Seq profiles and splicing sites of 33 genes that were associated with images, but not differentially expressed between tumor and normal tissue samples in the KIRC study, where dotted vertical lines denote location of splicing sites.

**Fig. S3.** Clusters of image associated genes in the KIRC study by $k$-means clustering algorithms.

**Fig. S4.**

**A.** Four genes TTC23, CPEB3, CAPN14 and PHKA1 with similar RNA-seq profiles and patterns of splicing sites in the ovarian cancer study formed a small cluster, where dotted vertical lines denote location of splicing sites.

**B.** Four genes CDCA2, TRAPPC11, PTPRG and ITGA10 with similar RNA-seq profiles and patterns of splicing sites in the ovarian cancer study formed a small cluster, where dotted vertical lines denote location of splicing sites.

**Supplementary Note 1**

**Two dimensional Functional Principal Component Analysis**

Consider a linear combination of functional values:

$$f = \iint_{S\,T} \beta(s,t)x(s,t)dsdt,$$

where $\beta(s,t)$ is a weight function and $x(s,t)$ is a centered random function. To capture the variations in the random functions, we chose weight function $\beta(s,t)$ to maximize the variance of $f$. By the formula for the variance of stochastic integral (1), we have

$$\text{var}(f) = \iiiint_{S\,T\,S\,T} \beta(s_1,t_1)R(s_1,t_1,s_2,t_2)\beta(s_2,t_2)ds_1 dt_1 ds_2 dt_2 , \quad (1)$$

where $R(s_1,t_1,s_2,t_2) = \text{cov}(x(s_1,t_1), x(s_2,t_2))$ is the covariance function of the genetic variant function $x(s,t)$. Since multiplying $\beta(t)$ by a constant will not change the maximizer of the variance $Var(f)$, we impose a constraint to make the solution unique:

$$\iint_{T\,T} \beta^2(s,t)dsdt = 1. \quad (2)$$

Therefore, to find the weight function, we seek to solve the following optimization problem:

$$\begin{aligned}
\max \quad & \iiiint_{S\,T\,S\,T} \beta(s_1,t_1)R(s_1,t_1,s_2,t_2)\beta(s_2,t_2)ds_1 dt_1 ds_2 dt_2 \\
\text{s.t.} \quad & \iint_{T\,T} \beta^2(s,t)dsdt = 1.
\end{aligned} \quad (3)$$

By the Lagrange multiplier, we reformulate the constrained optimization problem (3) into the following non-constrained optimization problem:

$$\max_{\beta} \frac{1}{2}\int_S\int_T\int_S\int_T \beta(s_1,t_1)R(s_1,t_1,s_2,t_2)\beta(s_2,t_2)ds_1dt_1ds_2dt_2 + \frac{1}{2}\lambda(1-\int_T\int_T \beta^2(s_1,t_1)ds_1dt_1), \quad (4)$$

where $\lambda$ is a parameter.

By variation calculus (2), we define the functional

$$J[\beta] = \frac{1}{2}\int_S\int_T\int_S\int_T \beta(s_1,t_1)R(s_1,t_1,s_2,t_2)\beta(s_2,t_2)ds_1dt_1ds_2dt_2 + \frac{1}{2}\lambda(1-\int_T\int_T \beta^2(s_1,t_1)ds_1dt_1).$$ Its first

variation is given by

$$\begin{aligned}
\delta J[h] &= \frac{d}{d\varepsilon}J[\beta(s,t)+\varepsilon h(s,t)] \\
&= \frac{d}{d\varepsilon}\{\frac{1}{2}\int_S\int_T\int_S\int_T \beta(s_1,t_1)R(s_1,t_1,s_2,t_2)\beta(s_2,t_2)ds_1dt_1ds_2dt_2 + \frac{1}{2}\lambda(1-\int_S\int_T \beta^2(s_1,t_1)ds_1dt_1)\}|_{\varepsilon=0} \\
&= \int_S\int_T\int_S\int_T h(s_1,t_1)R(s_1,t_1,s_2,t_2)\beta(s_2,t_2)ds_1dt_1ds_2dt_2 - \lambda\int_S\int_T \beta(s_1,t_1)h(s_1,t_1)ds_1dt_1 \\
&= \int_S\int_T[\int_S\int_T [R(s_1,t_1,s_2,t_2)\beta(s_2,t_2)ds_2dt_2 - \lambda\beta(s_1,t_1)]h(s_1,t_1)ds_1dt_1 \\
&= \int_S\int_T[\int_S\int_T R(s_1,t_1,s_2,t_2)\beta(s_2,t_2)ds_2dt_2 - \lambda\beta(s_1,t_1)]^2 ds_1dt_1 = 0,
\end{aligned}$$

which implies the following integral equation

$$\int_S\int_T R(s_1,t_1,s_2,t_2)\beta(s_2,t_2)ds_2dt_2 = \lambda\beta(s_1,t_1) \qquad (5)$$

for an appropriate eigenvalue $\lambda$. The left side of the integral equation (5) defines a two dimensional integral transform $R$ of the weight function $\beta$. Therefore, the integral transform of the covariance function $R(s_1,t_1,s_2,t_2)$ is referred to as the covariance operator $R$. The integral equation (5) can be rewritten as

$$R\beta = \lambda\beta, \qquad (6)$$

where $\beta(s_1,t_1,s_2,t_2)$ is an eigenfunction and referred to as a principal component function. Equation (6) is also referred to as a two dimensional eigenequation. Clearly, the eigenequation (6) looks the same as the eigenequation for the multivariate PCA if the covariance operator and eigenfunction are replaced by covariance matrix and eigenvector.

Since the number of function values is theoretically infinity, we may have an infinite number of eigenvalues. Provided the functions $X_i$ and $Y_i$ are not linearly dependent, there will be only $N-1$ nonzero eigenvalues, where $N$ is the total number of sampled individuals ($N = n_A + n_G$). Eigenfunctions satisfying the eigenequation are orthonormal (Ramsay and Silverman, 2005). In other words, equation (6) generates a set of principal component functions

$$R\beta_k = \lambda_k \beta_k, \qquad \text{with } \lambda_1 \geq \lambda_2 \geq \cdots$$

These principal component functions satisfy

(1) $\iint_{S\,T} \beta_k^2(s,t)\,dsdt = 1$ and

(2) $\iint_{S\,T} \beta_k(s,t)\beta_m(s,t)\,dsdt = 0,$ for all $m < k$.

The principal component function $\beta_1$ with the largest eigenvalue is referred to as the first principal component function, and the principal component function $\beta_2$ with the second largest eigenvalue is referred to as the second principal component function, and continues.

**Computations for the Principal Component Function and the Principal Component Score**

The eigenfunction is an integral function and difficult to solve in closed form. A general strategy for solving the eigenfunction problem in (5) is to convert the continuous eigen-analysis problem to an appropriate discrete eigen-analysis task (Ramsay and Silverman 2005). In this report, we use basis function expansion methods to achieve this conversion.

Let $\{\phi_j(t)\}$ be the series of Fourier functions. For each j, define $\omega_{2j-1} = \omega_{2j} = 2\pi j.$ We expand each genetic variant profile $x_i(s,t)$ as a linear combination of the basis function $\phi_j$:

$$x_i(s,t) = \sum_{k=1}^{K}\sum_{l=1}^{K} c_{kj}^{(i)} \phi_k(s)\phi_l(t). \qquad (7)$$

Let $C_i = [c_{11}^{(i)},...,c_{1K}^{(i)}, c_{21}^{(i)},...,c_{2K}^{(i)},...,c_{K1}^{(i)},...,c_{KK}^{(i)}]^T$ and $\phi(t) = [\phi_1(t),\cdots,\phi_K(t)]^T$. Then, equation (7) can be rewritten as

$$x_i(s,t) = C_i^T(\phi(s) \otimes \phi(t)),$$

where $\otimes$ denotes the Kronecker product of two matrices.

Define the vector-valued function $X(s,t) = [x_1(s,t),\cdots,x_N(s,t)]^T$. The joint expansion of all N random functions can be expressed as

$$X(s,t) = C(\phi(s) \otimes \phi(t)) \qquad (8)$$

where the matrix C is given by

$$C = \begin{bmatrix} C_1^T \\ \vdots \\ C_N^T \end{bmatrix}.$$

In matrix form we can express the variance-covariance function of the genetic variant profiles as

$$R(s_1,t_1,s_2,t_2) = \frac{1}{N} X^T(s_1,t_1)X(s_2,t_2) \\ = \frac{1}{N}[\phi^T(s_1) \otimes \phi^T(t_1)C^T C[\phi(s_2) \otimes \phi(t_2)]. \qquad (9)$$

Similarly, the eigenfunction $\beta(s,t)$ can be expanded as

$$\beta(s,t) = \sum_{j=1}^{K}\sum_{k=1}^{K} b_{jk}\phi_j(s)\phi_k(t) \text{ or}$$

$$\beta(s,t) = [\phi^T(s) \otimes \phi^T(t)]b, \qquad (10)$$

where $b = [b_{11},...,b_{1K},...,b_{K1},...,b_{KK}]^T$

Substituting expansions (9) and (10) of variance-covariance $R(s_1,t_1,s_2,t_2)$ and eigenfunction $\beta(s,t)$ into the functional eigenequation (5), we obtain

$$[\phi^T(s_1) \otimes \phi^T(t_1)]\frac{1}{N}C^T C b = \lambda[\phi^T(s_1) \otimes \phi^T(t_1)]b. \qquad (11)$$

Since equation (11) must hold for all $t$, we obtain the following eigenequation:

$$\frac{1}{N}C^T C b = \lambda b. \qquad (12)$$

Solving eigenequation (12), we obtain a set of orthonormal eigenvectors $b_j$. A set of orthonormal eigenfunctions is given by

$$\beta_j(s,t) = [\phi^T(s) \otimes \phi^T(t)]b_j, \; j = 1,...,J. \qquad (13)$$

The random functions $x_i(s,t)$ can be expanded in terms of eigenfunctions as

$$x_i(t,s) = \sum_{j=1}^{J} \xi_{ij}\beta_j(s,t), i = 1,...,N, \qquad (14)$$

where

$$\xi_{ij} = \iint_{S\,T} x_i(t,s)\beta_j(s,t)dsdt.$$

Max $\operatorname{Var}(\beta^T E[X - E(X)|Y]) = \beta^T \operatorname{cov}(E[X - E(X)|Y])\beta$
s.t. $\beta^T \operatorname{cov}(X,X)\beta = 1$

**Supplementary Note 2.**

**Multivariate Functional Regression Models for Quantitative Trait Analysis**

Assume that $n$ individuals are sampled. Let $y_{ik}, k = 1,2,...,K$ be $K$ trait values of the $i$-th individual. Consider a genomic region $[a, b]$. Let $x_i(t)$ be a RNA-seq profile of the $i$-th individual defined in the region $[a, b]$. Recall that a regression model for QTL analysis with the $k$-th trait and SNP data is defined as

$$y_{ik} = \mu_k + \sum_{j=1}^{J_1} x_{ij}\alpha_{kj} + \varepsilon_{ik} \qquad (1)$$

where $\mu_k$ is an overall mean of the $k$-th trait, $\alpha_{kj}$ is the main genetic additive effect of the $j$-th SNP in the genomic region for the $k$-th trait, $x_{ij}$ is an indicator variable for the genotypes at the $j$-th SNP, $\varepsilon_{ik}, k = 1,..,K$ are independent and identically distributed normal variables with mean of zero and covariance matrix $\Sigma$.

Similar to the multiple regression models for QTL analysis with SNP data and multiple quantitative traits, the functional regression model for a quantitative trait can with RNA-seq data can be defined as

$$y_{ik} = \alpha_{0k} + \int_T \alpha_k(t) x_i(t) dt + \varepsilon_{ik}, \qquad (2)$$

where $\alpha_{0k}$ is an overall mean, $\alpha_k(t)$ are a genetic additive effect of a putative QTLs located at the genomic positions $t$ for the $k$-th trait, $k = 1,...,K$, $x_i(t)$ is a genotype profile, $\varepsilon_{ik}$ are independent and identically distributed normal variables with mean of zero and covariance matrix $\Sigma$.

**Estimation of Additive Effects**

We assume that both phenotypes and genotype profiles are centered. The genotype profiles $x_i(t)$ are expanded in terms of the orthonormal basis function as:

$$x_i(t) = \sum_{j=1}^{\infty} \xi_{ij} \phi_j(t) \qquad (3)$$

where $\phi_j(t)$ are sequences of the orthonormal basis functions. The expansion coefficients $\xi_{ij}$ are estimated by

$$\xi_{ij} = \int_T x_i(t) \phi_j(t) dt \qquad (4)$$

In practice, numerical methods for the integral will be used to calculate the expansion coefficients.

Substituting equation (3) into equation (2), we obtain

$$\begin{aligned} y_{ik} &= \int_T \alpha_k(t) \sum_{j=1}^{\infty} \xi_{ij} \phi_j(t) dt + \varepsilon_i \\ &= \sum_{j=1}^{\infty} \xi_{ij} \int_T \alpha_k(t) \phi_j(t) dt + \varepsilon_{ik} \qquad (5) \\ &= \sum_{j=1}^{\infty} \xi_{ij} \alpha_{kj} + \varepsilon_{ik}, i = 1,...,n, k = 1,...,K, \end{aligned}$$

where $\alpha_{kj} = \int_T \alpha_k(t) \phi_j(t) dt$. The parameters $\alpha_{kj}$ are referred to as genetic additive effect scores for the $k$-th trait. These scores can also be viewed as the expansion coefficients of the genetic effect functions with respect to orthonormal basis functions:

$$\alpha_k(t) = \sum_j \alpha_{kj} \phi_j(t). \qquad (6)$$

Let

$$Y = [Y_1,...,Y_K] = \begin{bmatrix} Y_{11} & \cdots & Y_{1K} \\ \vdots & \ddots & \vdots \\ Y_{n1} & \cdots & Y_{nK} \end{bmatrix}, \xi = \begin{bmatrix} \xi_{11} & \cdots & \xi_{1J} \\ \vdots & \ddots & \vdots \\ \xi_{n1} & \cdots & \xi_{nJ} \end{bmatrix}, , \xi_i = \begin{bmatrix} \xi_{i1} \\ \vdots \\ \xi_{iJ} \end{bmatrix}, , ,$$

$$, \alpha_k = \begin{bmatrix} \alpha_{k1} \\ \vdots \\ \alpha_{kJ} \end{bmatrix}, \alpha = [\alpha_1,...,\alpha_K], \varepsilon = \begin{bmatrix} \varepsilon_{11} & \cdots & \varepsilon_{1K} \\ \cdots & \cdots & \cdots \\ \varepsilon_{n1} & \cdots & \varepsilon_{nK} \end{bmatrix}.$$

Then, equation (5) can be approximated by

$$Y = \xi\alpha + \varepsilon \qquad (7)$$

The standard least square estimators of $\alpha$ and the variance covariance matrix $\Sigma$ are given by

$$\hat{\alpha} = (\xi^T\xi)^{-1}\xi^T(Y-\bar{Y}), \qquad (8)$$

$$\hat{\Sigma} = \frac{1}{n}(Y-\xi\hat{\alpha})^T(Y-\xi\hat{\alpha}). \qquad (9)$$

Denote the matrix $(\xi^T\xi)^{-1}\xi^T$ by $A$. Then, the estimator of the parameter $\alpha$ is given by

$$\hat{\alpha} = A(Y-\bar{Y}). \qquad (10)$$

The vector of the matrix $\alpha$ can be written as

$$vec(\hat{\alpha}) = (A \otimes I)vec(Y-\bar{Y}). \qquad (11).$$

By the assumption of the variance matrix of $Y$, we obtain the variance matrix of $vec(Y)$:

$$var(vec(Y)) = \Sigma \otimes I. \qquad (12)$$

Thus, it follows from equations (11) and (12) that

$$\Lambda = \operatorname{var}(vec(\hat{\alpha})) = (I_k \otimes A)(\Sigma \otimes I_n)(I_k \otimes A^T)$$
$$= \Sigma \otimes (AA^T) \tag{13}$$

**Test Statistics**

An essential problem in QTL analysis or in integrative analysis of imaging and RNA-seq data is to test the **association of genomic region (or gene). Formally, we investigate the problem of** testing the following hypothesis:

$$\alpha_k(t) = 0, \forall t \in [a,b], k = 1,...,K,,$$

which is equivalent to testing the hypothesis:

$$H_0 : \alpha = 0.$$

Define the test statistic for testing the association of a genomic region with $K$ quantitative traits as

$$T = \hat{\alpha}^T \Lambda^{-1} \hat{\alpha}. \tag{14}$$

Let $r = \operatorname{rank}(\Lambda)$.

Then, under the null hypothesis $H_0 : \alpha = 0$, $T$ is asymptotically distributed as a central $\chi^2_{(KJ)}$ or $\chi^2_{(r)}$ distribution if $J$ components are taken in the expansion equation (3).

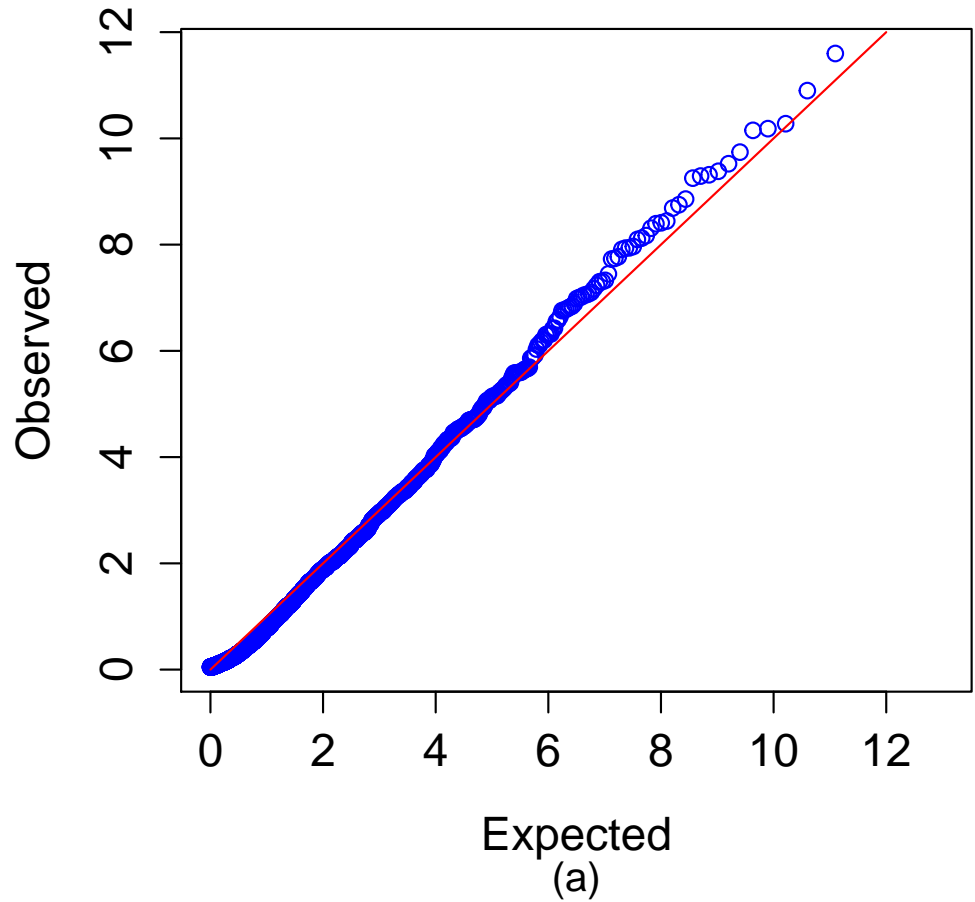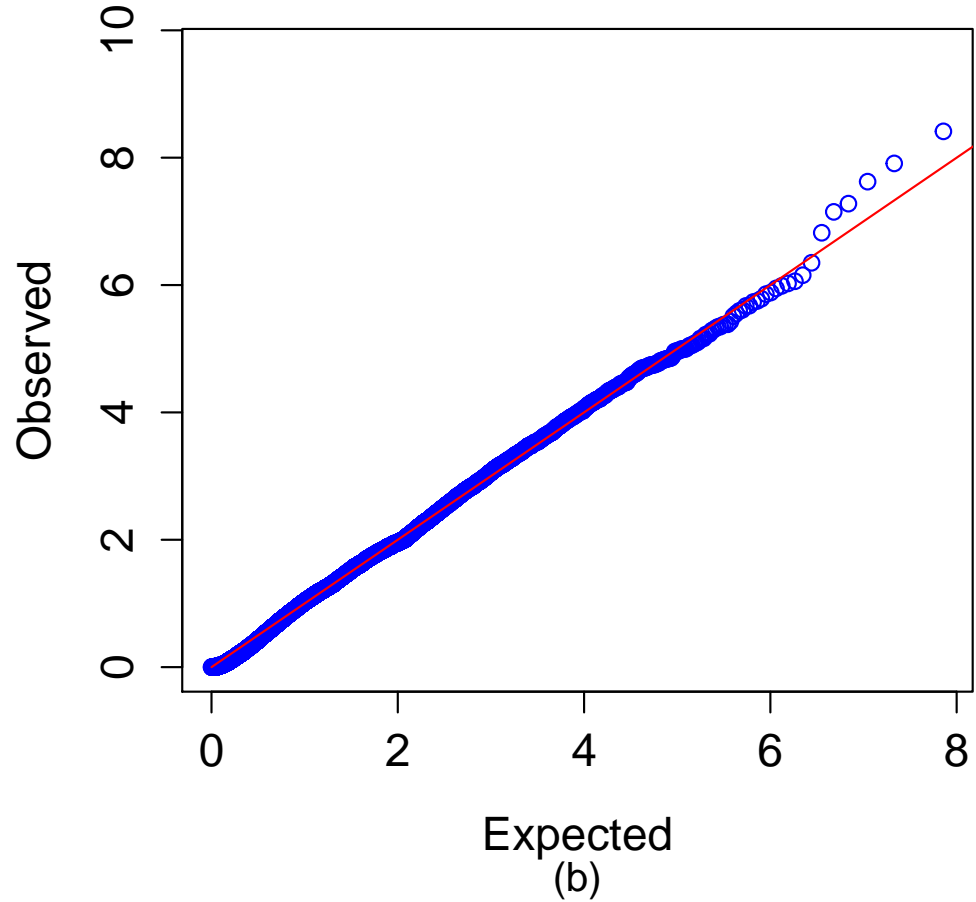

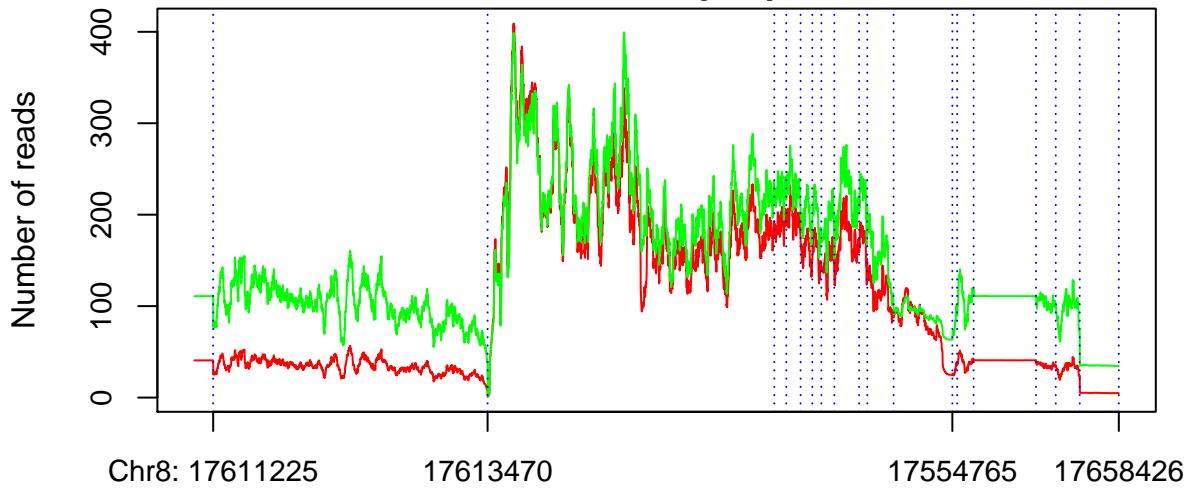

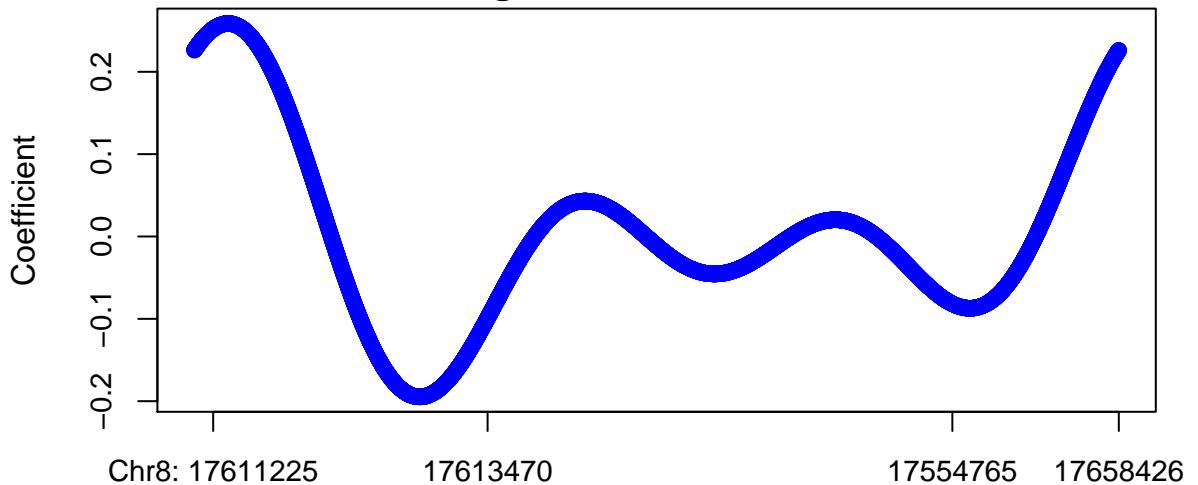

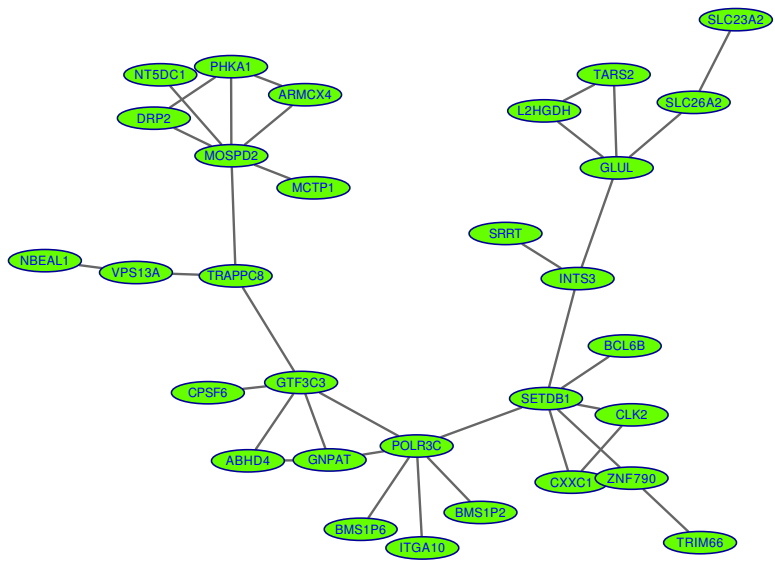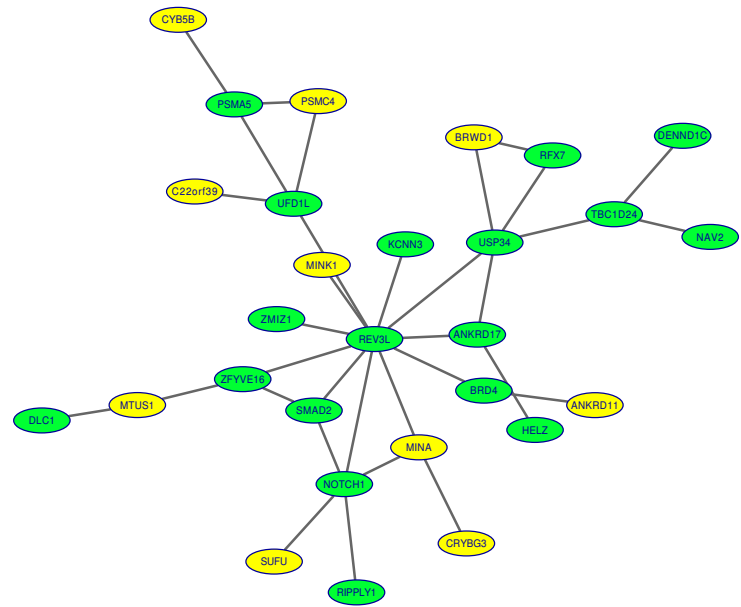

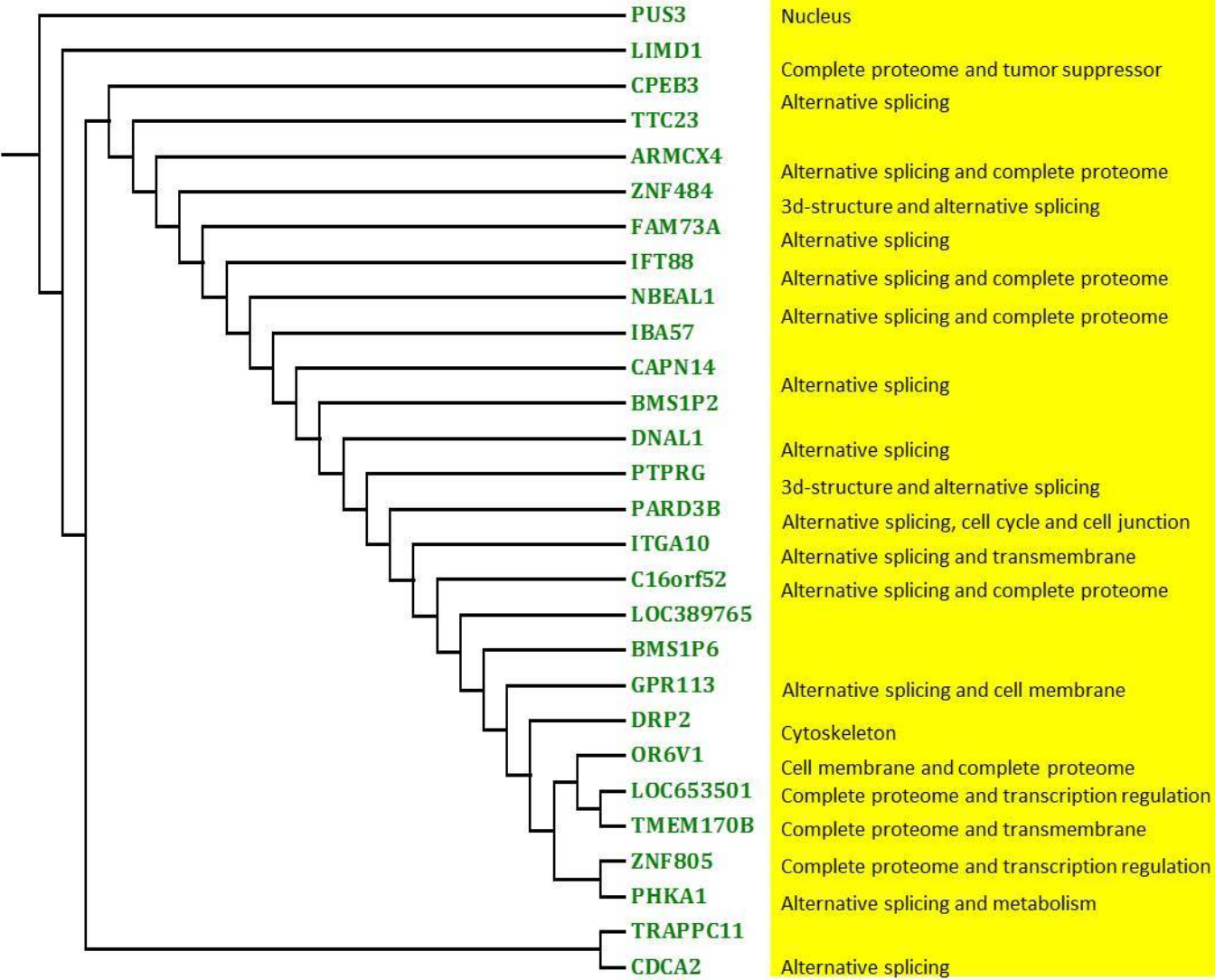

Table 1. P-values of three statistics for testing association of expression with images in ovarian cancer study.

| Gene | P-value | | |
|---|---|---|---|
| | MFLM_FPC | MFLM_Descriptor | MLM |
| ZNF805 | 2.31E-10 | 4.34E-01 | 9.33E-01 |
| LOC653501 | 3.86E-09 | 2.49E-02 | 9.04E-01 |
| TMEM170B | 1.23E-08 | 6.18E-03 | 9.11E-01 |
| DRP2 | 2.38E-08 | 9.76E-02 | 5.89E-01 |
| OR6V1 | 5.27E-08 | 2.07E-03 | 1.76E-01 |
| GPR113 | 7.09E-08 | 3.67E-07 | 5.70E-01 |
| ZNF484 | 4.47E-07 | 6.77E-03 | 9.34E-01 |
| DNAL1 | 7.00E-07 | 6.64E-03 | 8.59E-01 |
| ITGA10 | 8.72E-07 | 2.01E-01 | 6.41E-01 |
| NBEAL1 | 9.43E-07 | 7.67E-03 | 7.12E-01 |
| C16orf52 | 1.13E-06 | 2.10E-02 | 9.06E-01 |
| PHKA1 | 1.31E-06 | 2.80E-02 | 7.15E-01 |
| PTPRG | 1.39E-06 | 9.50E-01 | 6.58E-01 |
| IFT88 | 1.64E-06 | 1.09E-05 | 8.11E-01 |
| PARD3B | 1.78E-06 | 8.89E-01 | 4.49E-01 |
| LIMD1 | 2.11E-06 | 4.69E-01 | 8.71E-01 |
| FAM73A | 2.13E-06 | 3.97E-03 | 9.28E-01 |
| CAPN14 | 2.45E-06 | 1.78E-02 | 4.80E-01 |
| CPEB3 | 2.55E-06 | 2.62E-02 | 9.88E-01 |
| CDCA2 | 2.80E-06 | 9.73E-01 | 3.74E-01 |
| PUS3 | 3.08E-06 | 7.81E-01 | 9.22E-01 |

| Table 2. P-values of three statistics for testing association of expression with images in KIRC study. ||||||||
|---|---|---|---|---|---|---|---|
| Gene | P-value ||| Gene | P-value |||
| | MFLM (FPC) | MFLM (Descriptor) | MLM | | MFLM(FPC) | MFLM(Descriptor) | MLM |
| HELZ | 6.62E-16 | 6.08E-01 | 8.79E-01 | ZNF81 | 9.95E-08 | 2.06E-07 | 7.25E-01 |
| MARCH9 | 2.12E-15 | 1.02E-06 | 7.58E-01 | GAB2 | 1.04E-07 | 1.34E-02 | 6.38E-01 |
| MSH5-SAPCD1 | 8.98E-13 | 9.79E-03 | NA* | MMP24-AS1 | 1.29E-07 | 3.26E-06 | NA |
| SLC2A12 | 2.52E-12 | 9.94E-03 | 2.76E-08 | LOC647859 | 1.43E-07 | 8.56E-02 | 1.23E-03 |
| BRWD1 | 1.26E-11 | 5.61E-03 | 9.54E-01 | C2orf68 | 1.49E-07 | 4.83E-03 | 7.84E-01 |
| RFX7 | 5.29E-11 | 1.00E+00 | 9.58E-01 | SDR39U1 | 1.57E-07 | 8.83E-04 | 5.88E-01 |
| C22orf39 | 6.55E-11 | 1.29E-03 | 5.77E-01 | ZRANB3 | 1.66E-07 | 1.03E-03 | 9.59E-01 |
| NSD1 | 7.06E-11 | 1.67E-02 | 9.74E-01 | PSMC4 | 1.71E-07 | 1.39E-02 | 8.87E-01 |
| RTF1 | 1.82E-10 | 9.49E-01 | 8.58E-01 | FLJ12825 | 1.74E-07 | 1.39E-04 | 7.08E-01 |
| MBD5 | 3.00E-10 | 1.08E-04 | 9.33E-01 | ARHGEF11 | 2.26E-07 | 8.24E-03 | 8.55E-01 |
| ZSCAN16-AS1 | 4.16E-10 | 6.08E-02 | NA | LOC100289019 | 2.61E-07 | 8.50E-04 | NA |
| SESN1 | 4.84E-10 | 3.42E-01 | 6.71E-01 | SUFU | 2.79E-07 | 1.99E-01 | 5.84E-01 |
| ITGA9 | 5.12E-10 | 2.11E-02 | 9.52E-01 | ZNF555 | 3.75E-07 | 2.16E-02 | 3.75E-01 |
| PPM1K | 5.60E-10 | 1.48E-01 | 1.11E-04 | KHNYN | 3.85E-07 | 1.54E-01 | 4.62E-01 |
| USP42 | 1.39E-09 | 9.79E-01 | 9.06E-01 | ANKRD11 | 4.80E-07 | 1.00E+00 | 8.92E-01 |
| FAM47E-STBD1 | 1.77E-09 | 1.11E-02 | NA | BOLA2 | 4.82E-07 | 9.88E-02 | 8.33E-01 |
| ZNF710 | 2.05E-09 | 1.22E-01 | 9.82E-01 | BOLA2B | 4.82E-07 | 9.88E-02 | NA |
| TECPR2 | 3.59E-09 | 9.53E-04 | 5.63E-01 | SAPCD1 | 4.97E-07 | 4.24E-01 | NA |
| RASSF8-AS1 | 3.88E-09 | 3.08E-03 | NA | SLC9A4 | 6.26E-07 | 2.27E-02 | 1.87E-01 |
| CCDC93 | 4.04E-09 | 1.00E+00 | 9.45E-01 | CRYBG3 | 6.30E-07 | 5.18E-03 | 5.59E-02 |
| NAV2 | 4.90E-09 | 4.11E-02 | 1.17E-01 | SLC15A2 | 6.78E-07 | 1.18E-04 | 3.63E-05 |
| CYB5B | 6.75E-09 | 5.52E-04 | 7.11E-01 | BRD4 | 7.77E-07 | 5.46E-01 | 9.85E-01 |
| ANKRD17 | 7.55E-09 | 1.00E+00 | 4.47E-01 | ATP6V1C2 | 7.80E-07 | 2.86E-03 | 1.88E-01 |
| CCDC181 | 7.98E-09 | 5.24E-03 | NA | SMAD2 | 9.23E-07 | 9.38E-01 | 7.52E-01 |
| SPHK2 | 1.08E-08 | 3.26E-03 | 1.83E-02 | ST3GAL6 | 1.19E-06 | 5.01E-01 | 2.16E-01 |

| Gene | p-value | | | Gene | p-value | | |
|---|---|---|---|---|---|---|---|
| KCNN3 | 1.15E-08 | 9.47E-01 | 9.17E-01 | ZMIZ1 | 1.26E-06 | 3.75E-01 | 9.46E-01 |
| ZFYVE16 | 1.16E-08 | 1.98E-02 | 7.65E-01 | USP34 | 1.36E-06 | 1.74E-03 | 8.24E-01 |
| CMTM1 | 1.23E-08 | 9.99E-01 | 3.66E-01 | RALGAPA2 | 1.38E-06 | 3.48E-03 | 1.57E-01 |
| LINC00875 | 1.69E-08 | 1.00E+00 | NA | FRMD4A | 2.03E-06 | 5.14E-03 | 7.95E-01 |
| NOTCH1 | 1.81E-08 | 1.96E-02 | 1.80E-01 | PSMA5 | 2.12E-06 | 7.81E-02 | 9.35E-01 |
| BLZF1 | 1.87E-08 | 2.05E-03 | 8.94E-01 | RIPPLY1 | 2.18E-06 | 1.00E+00 | 3.89E-06 |
| CHD2 | 3.55E-08 | 1.28E-01 | 9.47E-01 | ERCC6 | 2.20E-06 | 1.21E-01 | 7.01E-01 |
| MTUS1 | 4.71E-08 | 8.63E-02 | 2.18E-01 | MINK1 | 2.29E-06 | 1.19E-02 | 9.14E-01 |
| REV3L | 4.96E-08 | 2.28E-02 | 9.59E-01 | DIP2C | 2.38E-06 | 2.51E-03 | 9.13E-01 |
| LRIG2 | 5.00E-08 | 6.78E-03 | 8.40E-01 | PHLDB2 | 2.51E-06 | 3.45E-03 | 6.34E-01 |
| DENND1C | 5.83E-08 | 9.41E-01 | 2.40E-01 | TBC1D24 | 2.54E-06 | 7.79E-01 | 2.13E-01 |
| TMEM50B | 6.77E-08 | 4.16E-03 | 8.31E-02 | APBA3 | 2.55E-06 | 7.20E-02 | 1.29E-01 |
| CELF1 | 7.92E-08 | 1.00E+00 | 9.13E-01 | TRAK1 | 2.60E-06 | 7.67E-01 | 1.53E-01 |
| ZSCAN20 | 8.41E-08 | 1.23E-06 | 9.37E-01 | DLC1 | 2.60E-06 | 7.23E-02 | 8.35E-01 |
| MINA | 8.76E-08 | 1.61E-03 | 4.43E-03 | NISCH | 2.62E-06 | 1.00E+00 | 2.28E-01 |
| C5AR2 | 9.01E-08 | 4.97E-02 | NA | CPEB3 | 2.64E-06 | 9.82E-01 | 4.91E-03 |
| SSH2 | 9.68E-08 | 1.00E-02 | 3.00E-01 | UFD1L | 2.94E-06 | 3.03E-03 | 7.78E-01 |

[*] NA: Expression (level 3) data were not available.

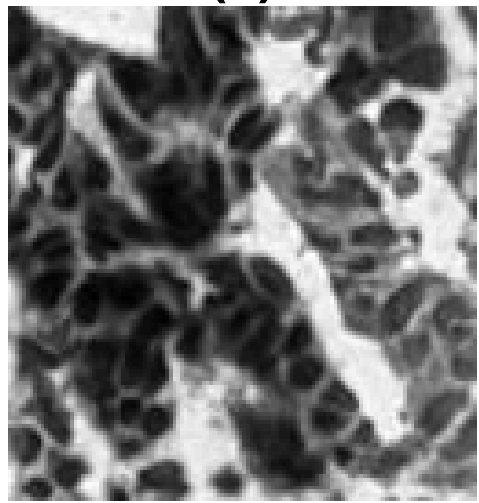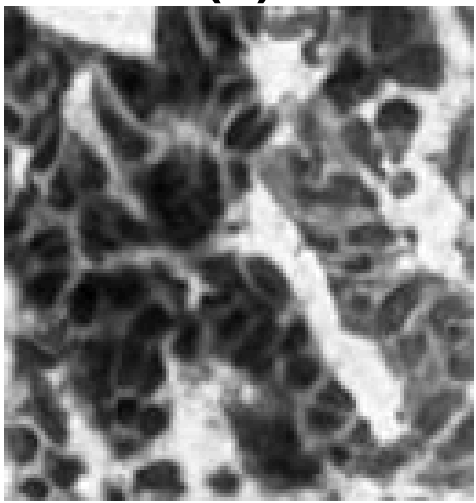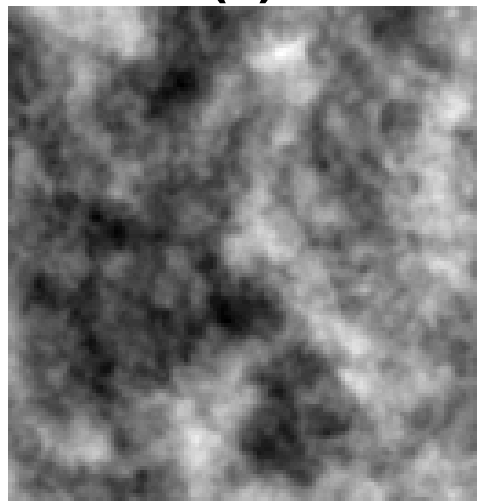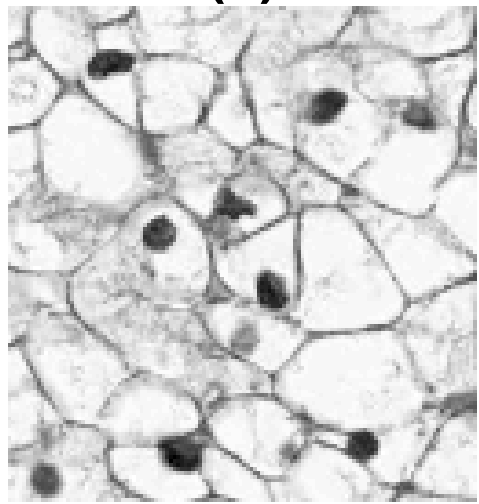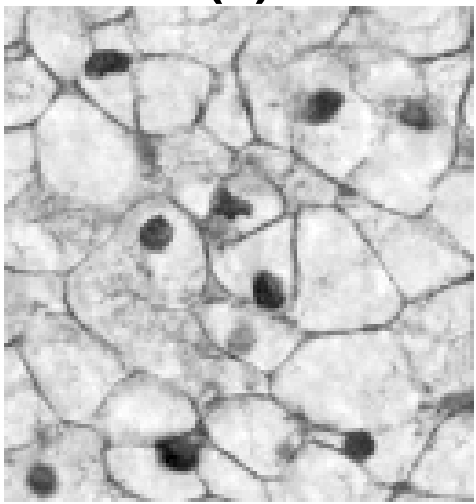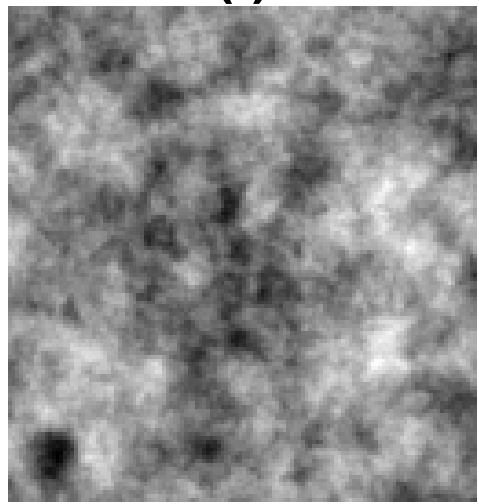

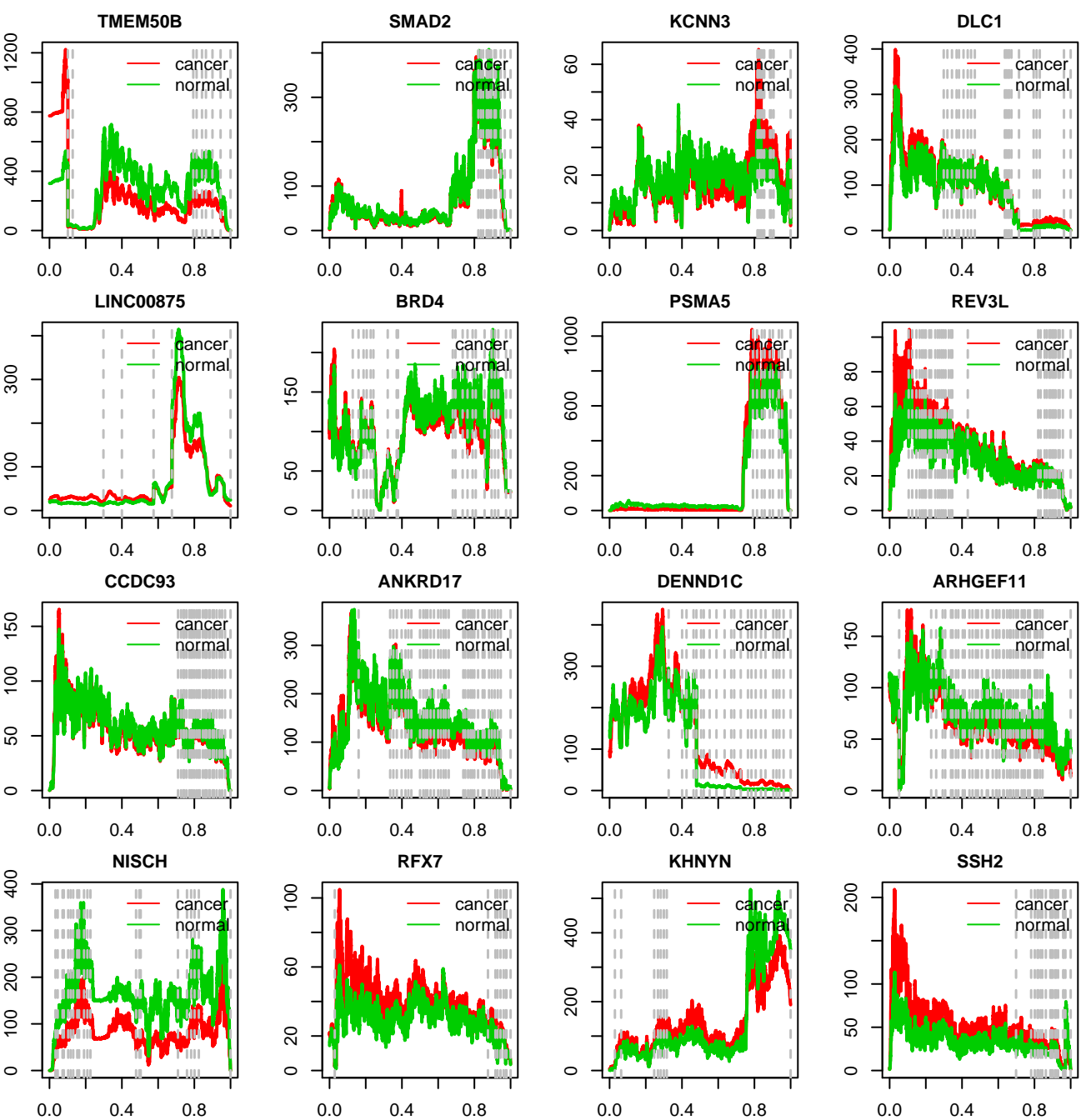

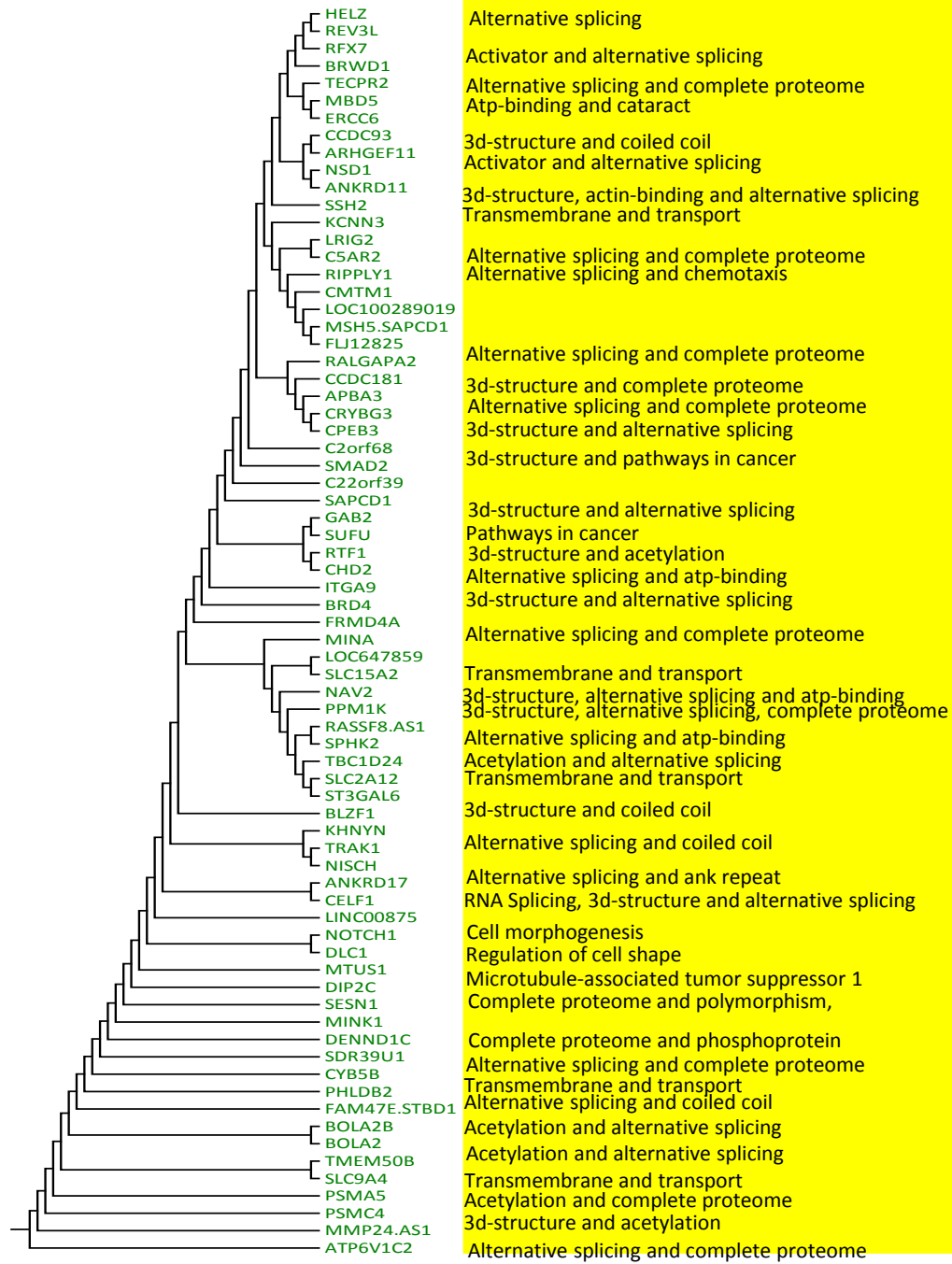

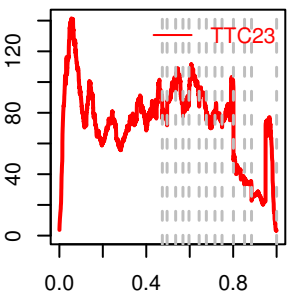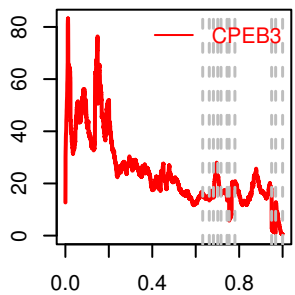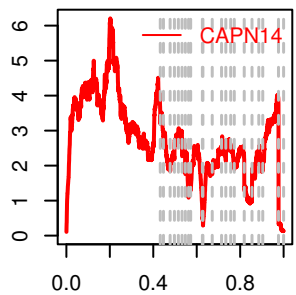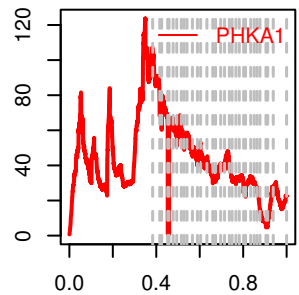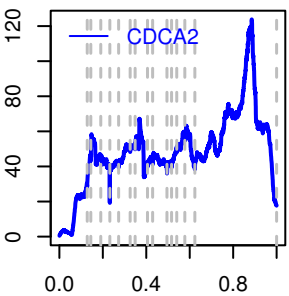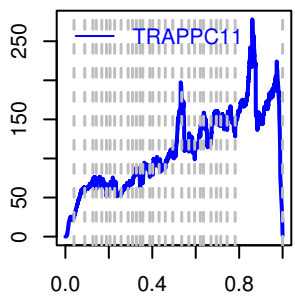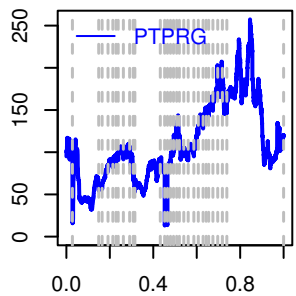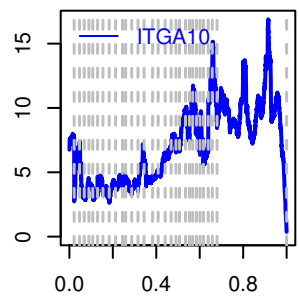

Table S1. P-values of the test for association of gene expressions with imaging and differential expressions between tumor and normal tissue samples in the KIRC study.

| Gene | P-values | | Gene | P-values | |
|---|---|---|---|---|---|
| | MFLM_FPC | Differential Expression | | MFLM_FPC | Differential Expression |
| HELZ | 6.62E-16 | 8.65E-08 | ZNF81 | 9.95E-08 | NA |
| MARCH9 | 2.12E-15 | NA | GAB2 | 1.04E-07 | 9.16E-06 |
| MSH5.SAPCD1 | 8.98E-13 | 7.13E-12 | MMP24.AS1 | 1.29E-07 | 6.58E-01 |
| SLC2A12 | 2.52E-12 | <1.0 E-27 | LOC647859 | 1.43E-07 | 3.00E-02 |
| BRWD1 | 1.26E-11 | 9.17E-08 | C2orf68 | 1.49E-07 | 3.27E-06 |
| RFX7 | 5.29E-11 | 1.48E-01 | SDR39U1 | 1.57E-07 | 6.66E-16 |
| C22orf39 | 6.55E-11 | 2.52E-11 | ZRANB3 | 1.66E-07 | NA |
| NSD1 | 7.06E-11 | 4.44E-16 | PSMC4 | 1.71E-07 | 5.11E-11 |
| RTF1 | 1.82E-10 | <1.0 E-27 | FLJ12825 | 1.74E-07 | NA |
| MBD5 | 3.00E-10 | 3.47E-13 | ARHGEF11 | 2.26E-07 | 6.50E-02 |
| ZSCAN16.AS1 | 4.16E-10 | NA | LOC100289019 | 2.61E-07 | 9.86E-02 |
| SESN1 | 4.84E-10 | 4.22E-02 | SUFU | 2.79E-07 | 4.45E-07 |
| ITGA9 | 5.12E-10 | 4.13E-08 | ZNF555 | 3.75E-07 | NA |
| PPM1K | 5.60E-10 | 1.11E-16 | KHNYN | 3.85E-07 | 2.44E-01 |
| USP42 | 1.39E-09 | NA | ANKRD11 | 4.80E-07 | 7.49E-06 |
| FAM47E.STBD1 | 1.77E-09 | 1.54E-02 | BOLA2B | 4.82E-07 | 4.65E-06 |
| ZNF710 | 2.05E-09 | NA | BOLA2 | 4.82E-07 | 6.96E-06 |
| TECPR2 | 3.59E-09 | 2.38E-11 | SAPCD1 | 4.97E-07 | 6.51E-09 |
| RASSF8.AS1 | 3.88E-09 | 7.97E-01 | SLC9A4 | 6.26E-07 | NA |
| CCDC93 | 4.04E-09 | 4.13E-03 | CRYBG3 | 6.30E-07 | 1.58E-06 |
| NAV2 | 4.90E-09 | 3.51E-01 | SLC15A2 | 6.78E-07 | 8.05E-11 |
| CYB5B | 6.75E-09 | 2.25E-08 | BRD4 | 7.77E-07 | 8.09E-04 |
| ANKRD17 | 7.55E-09 | 8.11E-03 | ATP6V1C2 | 7.80E-07 | 1.05E-07 |
| CCDC181 | 7.98E-09 | NA | SMAD2 | 9.23E-07 | 5.33E-05 |
| SPHK2 | 1.08E-08 | 3.84E-07 | ST3GAL6 | 1.19E-06 | 1.53E-08 |
| KCNN3 | 1.15E-08 | 1.57E-04 | ZMIZ1 | 1.26E-06 | NA |
| ZFYVE16 | 1.16E-08 | NA | USP34 | 1.36E-06 | NA |
| CMTM1 | 1.23E-08 | 2.64E-03 | RALGAPA2 | 1.38E-06 | 7.06E-13 |
| LINC00875 | 1.69E-08 | 7.54E-04 | FRMD4A | 2.03E-06 | 0.00E+00 |
| NOTCH1 | 1.81E-08 | 3.71E-02 | PSMA5 | 2.12E-06 | 1.95E-03 |
| BLZF1 | 1.87E-08 | 4.03E-12 | RIPPLY1 | 2.18E-06 | NA |
| CHD2 | 3.55E-08 | 2.57E-08 | ERCC6 | 2.20E-06 | 0.00E+00 |
| MTUS1 | 4.71E-08 | 9.48E-11 | MINK1 | 2.29E-06 | 0.00E+00 |
| REV3L | 4.96E-08 | 2.12E-03 | DIP2C | 2.38E-06 | 1.32E-14 |
| LRIG2 | 5.00E-08 | 1.25E-13 | PHLDB2 | 2.51E-06 | 6.66E-16 |
| DENND1C | 5.83E-08 | 1.20E-02 | TBC1D24 | 2.54E-06 | 8.88E-01 |
| TMEM50B | 6.77E-08 | 1.08E-05 | APBA3 | 2.55E-06 | 5.83E-14 |
| CELF1 | 7.92E-08 | 3.99E-10 | DLC1 | 2.60E-06 | 2.20E-04 |
| ZSCAN20 | 8.41E-08 | NA | TRAK1 | 2.60E-06 | 5.01E-02 |
| MINA | 8.76E-08 | < 1.0 E-27 | NISCH | 2.62E-06 | 1.03E-01 |
| C5AR2 | 9.01E-08 | <1.0 E-27 | CPEB3 | 2.64E-06 | 5.64E-01 |
| SSH2 | 9.68E-08 | 3.85E-01 | UFD1L | 2.94E-06 | NA |

NA: Test for differential expression was not conducted due to technique problems.